\documentclass[aps,prb,floatfix,superscriptaddress,twocolumn,footinbib]{revtex4-1}

\usepackage{enumerate}

\usepackage{amssymb}
\usepackage{amsmath}
\usepackage{amsfonts}
\usepackage{appendix}
\usepackage{bm}
\usepackage{graphicx}
\usepackage{epsfig}
\usepackage{epstopdf}
\usepackage{balance}
\usepackage[dvipsnames]{xcolor}
\usepackage{calc}
\usepackage{natbib}
\usepackage{multirow}
\usepackage[colorlinks,
            linkcolor=blue,
            anchorcolor=blue,
            citecolor=blue,
            urlcolor=blue]{hyperref}
\usepackage{lipsum}
\usepackage{enumerate}

\begin{document}

\title{Moir\'{e} Band Structures of the Double twisted Few Layer Graphene}
\author{Miao Liang}
\affiliation{School of Physics and Wuhan National High Magnetic Field Center,
Huazhong University of Science and Technology, Wuhan 430074,  China}

\author{Meng-Meng Xiao}
\affiliation{School of Physics and Wuhan National High Magnetic Field Center,
Huazhong University of Science and Technology, Wuhan 430074,  China}

\author{Zhen Ma}
\email{mazhen@hust.edu.cn}
\affiliation{School of Physics and Wuhan National High Magnetic Field Center,
Huazhong University of Science and Technology, Wuhan 430074,  China}

\author{Jin-Hua Gao}
\email{jinhua@hust.edu.cn}
\affiliation{School of Physics and Wuhan National High Magnetic Field Center,
Huazhong University of Science and Technology, Wuhan 430074,  China}

\begin{abstract}
Very recently, unconventional superconductivity has been observed in the double twisted trilayer graphene (TLG), where three monolayer graphene (MLG) are stacked on top of each other with two twist angles [J. M. Park, \textit{et al.}, \href{https://doi.org/10.1038/s41586-021-03192-0}{Nature \textbf{590}, 249 (2021)}; Z. Hao, \textit{et al.}, \href{https://www.science.org/doi/abs/10.1126/science.abg0399}{Science \textbf{371}, 1133 (2021)}; X. Zhang, \textit{et al.}, \href{https://link.aps.org/doi/10.1103/PhysRevLett.127.166802}{Phys. Rev. Lett.\textbf{127}, 166802 (2021)}]. When some of MLGs in the double twisted TLG are replaced by bilayer graphene (BLG), we get a new family of double twisted moir\'{e} heterostructure, namely double twisted few layer graphene (DTFLG). In this work, we theoretically investigate the moir\'{e} band structures of the DTFLGs with diverse arrangements of MLG and BLG. We find that, depending on the relative rotation direction of the two twist angles (alternate or chiral twist) and the middle van der Waals (vdW) layer (MLG or BLG), a general (X+Y+Z)-DTFLG can be classified into four categories, \textit{i.e.}~(X+1+Z)-ATFLG, (X+2+Z)-ATFLG, (X+1+Z)-CTFLG and (X+2+Z)-CTFLG, each of which has its own unique band structure. Here, X, Y, Z denote the three vdW layers, \textit{i.e.}~MLG or BLG. Interestingly, the (X+1+Z)-ATFLGs have a pair of perfect flat bands at the magic angle about $1.54^\circ$ coexisting with a pair of linear or parabolic bands, which is quite like the double twisted TLG. Meanwhile, when the twist angle is smaller than a ``magic angle'' $1.70^\circ$, the (X+2+Z)-CTFLGs can have two isolated narrow bands at $E_f$ with band width less than $5$ meV.  The influence of electric field and the topological features of the moir\'{e} bands have been studied as well. Our work indicates that the DTFLGs, especially the (X+1+Z)-ATFLG and (X+2+Z)-CTFLG, are promising platforms to study the moir\'{e} flat band induced novel correlation and topological effects.

\end{abstract}

\cite{}
\maketitle
\section{INTRODUCTION}
The discovery of unconventional superconductivity and broken-symmetry states in alternating twisted trilayer graphene (ATTLG) has drawn great research interest very recently\cite{park_tunable_2021,0Electric,cao_pauli-limit_2021,PhysRevResearch.2.033357,PhysRevB.104.L121116,Scammell2021}. It is of special importance because that the ATTLG gives a second definite example of moir\'{e} superconductor in addition to the celebrated twisted bilayer graphene (TBG)\cite{cao2018b,cao2018a,Jiang201991,Yankowitz2019,lau2019,Kere2019,koshino2018,kangjian2018,Guo2018,wu2018,liu2018,zhangyi2020,PhysRevResearch.3.033260,lu_superconductors_2019,stepanov_untying_2020,PhysRevB.98.220504,PhysRevLett.122.257002,Guinea13174,helin121406}.  

Different from the TBG, the ATTLG is a kind of double twisted moir\'{e} heterostructure\cite{Khalaf2019,lixiaotrilayer,Carr9b04979,Shin2021,2021Lattice,PhysRevB.104.035139}, which has two twist angles (with opposite rotation directions) instead of one as that in the TBG. Very interestingly, novel superconductivity has also been reported in another kind of double twisted trilayer graphene (TLG) with different twist manner\cite{PhysRevLett.127.166802}, \textit{i.e.}~chirally twisted TLG (CTTLG)\cite{Mora2019,zhu2020,PhysRevB.101.224107,Fengcheng2020,cea2019twists,helin97.035440}, where the two twist angles have the same rotation direction. These intriguing experimental findings imply that the double twisted moir\'{e} heterostructures should be  promising platforms to study the exotic correlation phases, \textit{e.g.}~unconventional superconductivity, based on moir\'{e} flat bands. However, the moir\'{e} band structures of other double twisted moir\'{e} materials have been paid little attention so far\cite{2021Doubled,Tritsaris2020,Wang2019,2020Double3,leconte2020}, though the band structures of the ATTLG and CTTLG were theoretically predicted two years ago.

 \begin{figure}[ht]
\centering
\includegraphics[width=8cm]{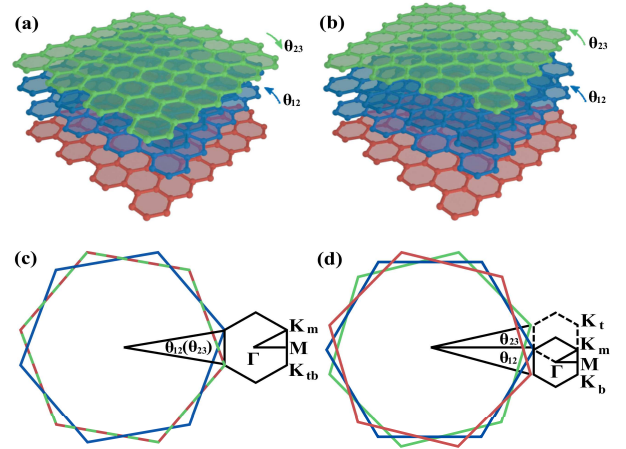}
\caption{Schematic of the the DTFLG.  (a) is the $(1+2+1)$-ATFLG where $\theta_{12}$ and $\theta_{23}$ have the opposite rotation direction.  (b) is the $(1+2+1)$-CTFLG where $\theta_{12}$ and $\theta_{23}$ have the same rotation direction.   (c) is the BZ of ATFLG and (d) is the BZ of CTFLG.  Red, blue and green represent the bottom, middle and top vdW layer, respectively.
}
\label{fig1}  
\end{figure}

\begin{figure*}[ht]
\centering
\includegraphics[width=14cm]{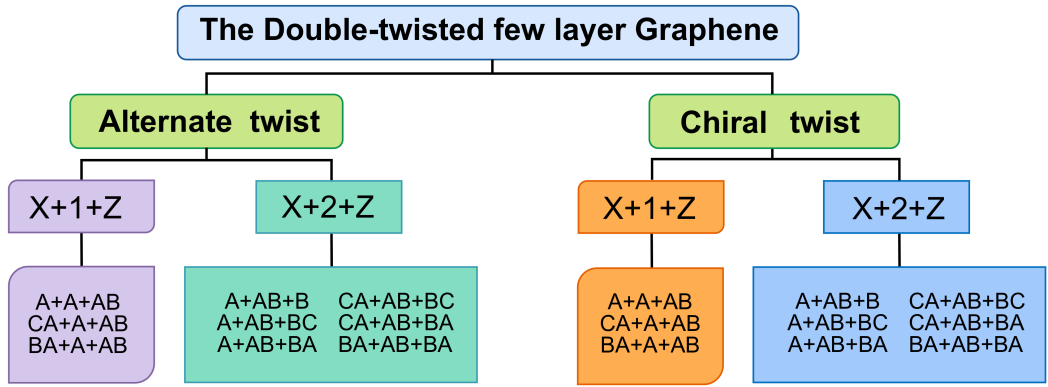}
\caption{Classification of the DTFLGs, according to their different kinds of moir\'{e} band structures.  All the possible arrangements of the MLG and BLG for each category of DTFLG are given as well.  
}
\label{fig2}
\end{figure*}

Generally speaking, a double twisted moir\'{e} heterostructure consists of three vdW layers, which are stacked on top of each other with two twist angles, as illustrated in Fig.~\ref{fig1}. According to the relative rotation direction of the two twist angles,  the double twisted moir\'{e} heterostructures fall into two categories:
(1) Alternating twisted one, where two twist angles have opposite rotation direction; (2) Chirally twisted one, where two twist angles have the same rotation direction. Different twist manner will give rise to fundamentally different moir\'{e} band structures.  The double twisted TLG  is a typical example. The ATTLG has a pair of flat bands at $E_f$ coexisting with a pair of linear bands (Dirac cone), where the magic angle here is about $1.54^\circ$ larger than the  magic angle $1.05^\circ$ of the TBG\cite{Khalaf2019,lixiaotrilayer, Carr9b04979,Shin2021}. In contrast, the CTTLG  is a perfect metal gapless at all energies\cite{Mora2019}. 
Meanwhile, using different kinds of vdW layers also can give rise to various moir\'{e} band structures\cite{jinhuaGao2020,2021Doubled,zhang2020_1,cheng2021}. In double twisted TLG, the three vdW layers are all monolayer graphene (MLG). Actually, if graphene multilayers are used as building blocks to construct  double twisted moir\'{e} heterostructure, namely \textit{double twisted few layer graphene} (DTFLG),  more interesting moir\'{e} flat band structures can be achieved.
For example, our recent work\cite{2021Doubled} predicts that,  once a ABA-stacked TLG is used to replace the the middle vdW layer of the double twisted TLG, it can produce doubled moir\'{e} flat bands, \textit{i.e.}, four degenerate moir\'{e} flat bands at $E_f$. It directly doubles the density of states (DOS) at $E_f$, and thus will greatly enhance the correlation effects.

In this work, we  study  the simplest kind of DTFLG other than the double twisted TLG, in which the three vdW layers are either MLG or bilayer graphene (BLG).
 According to whether it is alternating twisted or chirally twisted, a general DTFLG can be denoted as (X+Y+Z)-ATFLG or (X+Y+Z)-CTFLG,  where X, Y and Z represent the top, middle and bottom vdW layers, respectively. Here, we require that the X, Y, Z to be  either MLG or BLG, and at least one of them is BLG in order to be distinguished from the double twisted TLG.  For example, Fig.~\ref{fig1} (a) is the schematic of the (1+2+1)-ATFLG with Y=BLG (or denoted as 2) and X=Z=MLG (or denoted as 1) , while Fig.~\ref{fig1} (b) is for the (1+2+1)-CTFLG.   Unless specified otherwise, DTFLG hereafter always refers to the one with either MLG or BLG as its vdW layers as defined above. 

As well known, BLG is the simplest graphene multilayer in nature. It has two bands with $k^2$ dispersion touching at the Dirac points, and a perpendicular electric field can open a gap at $E_f$\cite{McCann_2013}.  In comparison, MLG has two linear bands forming a Dirac cone, which can not be gapped by an external electric field. So, whichever you choose (MLG or BLG),  each vdW layer always offers two bands near $E_f$ which are coupled by the same moir\'{e} interlayer hopping.  A good analogy in single twist devices is the relation among the TBG, twisted double-bilayer graphene (TDBG) \cite{lee_theory_2019,PhysRevB.99.235406,cao2019,2019arXiv190308130L,wu2019,Samajdar_2021,lee2019,zhangguangyu2019,jeil2019,zhangguangyu2019} and twisted TLG (\textit{i.e.}~twisted monolayer-bilayer graphene)\cite{MA202118,shi2020,young2020,chenshaowen2020,brey2013,xu2021tunable,Minhao2021,PhysRevB.102.035411,PhysRevResearch.2.033150}.  It is known that, once  the MLGs in the TBG are replaced by BLG, similar but different moir\'{e} flat band structures are achieved\cite{ZhangYaHuiprB.99.075127, PhysRevB.99.235406, MA202118}. 
For the double twist situation, it is interesting to see what kinds of moir\'{e} flat bands we can get if one or several MLGs in double twisted TLG are replaced by BLGs, and it is the main motivation of this paper.

Here, based on the continuous model method\cite{mac2011,prl2007,prb2012,KoshinoprB.87.205404,Koshino_2015}, we systematically calculate the moir\'{e} band structures of all the possible DTFLGs. Our numerical calculations indicate that the DTFLGs with  diverse arrangements of the vdW layers  indeed have  various exotic moir\'{e} band structures, quite dissimilar to the well known double twisted TLGs.  For the case of alternate twist, we find that the dominate factor is the middle vdW layer. That the middle vdW layer is MGL or BLG will produce completely different kinds of  moir\'{e} band structures: 
\begin{enumerate}
    \item (X+1+Z)-ATFLG. A pair of perfect flat bands at the magic angle $1.54^\circ$. The magic angle here is the same as that of the ATTLG.
    If the top and bottom vdW layers are both BLG, the flat bands here will coexist with a pair of parabolic bands touching at $E_f$.  Otherwise, it has a pair of linear bands coexisting with the flat bands, exactly the same as the ATTLG. In all cases, a perpendicular electric field can isolate the two flat bands, which then have nonzero valley Chern number.     
    
    \item (X+2+Z)-ATFLG.  Two or four dispersive bands at $E_f$, where the band width of the inner two can be greatly reduced as the twist angle decreasing. Note that there is no perfect flat bands at any twist angle here. Specifically,  if it has a AB/BA moir\'{e} interface, there are four dispersive bands degenerate at one Dirac point. Otherwise, only two bands appears at $E_f$.   The two narrow bands most close to $E_f$ can be isolated by an electric field, which  have  nonzero valley Chern number as well.  
\end{enumerate}
For the case of chiral twist, the middle  vdW layer is still an important factor  for the band structures. But the differences of the moir\'{e} bands between the (X+1+Z)-CTFLG and (X+2+Z)-CTFLG are not as obvious as that in the alternate twist case. 
The basic features of the moir\'{e} bands in this situation are:
\begin{enumerate}
    \item Always two dispersive bands  at $E_f$,   the band width of which can be greatly reduced as the twist angle decreasing. No perfect flat bands at any twist angle.  Except the case of (CA+A+AB)-CTFLG, such two narrow bands are  isolated from other high energy bands, which is distinct from the known CTTLG. 
    
    \item At the same twist angle, the band width of the two moir\'{e} bands at $E_f$ in (X+2+Z)-CTFLG is much smaller than that in (X+1+Z)-CTFLG. When the twist angle is  smaller than a ``magic angle'' about $1.70^\circ$, the (X+2+Z)-CTFLG can have two narrow bands with band width less than $5$ meV.  Meanwhile, in (X+2+Z)-CTFLG, two additional narrow bands will come very close to $E_f$ when $\theta<0.82^\circ$, namely, four narrow bands appear near $E_f$.  It thus will greatly enhance the density of states (DOS) at $E_f$ further.
    
    \item The two narrow bands at $E_f$ are degenerate at the Dirac points of the vdW layers. Electric field can lift all the degeneracy at the Dirac points belonging to the BLG (as a vdW layer). The narrow band isolated by the electric field has nonzero valley Chern number. 
    \end{enumerate}
    
So, the DTFLGs can be categorized into four basic types, as shown Fig.~\ref{fig2}, each of which has its own unique moir\'{e} band structures. According to the our numerical results, we suggest that the (X+1+Z)-ATFLG and (X+2+Z)-CTFLG are  of most interest to further experiments, which are most likely to host prominent correlation effects. The (X+1+Z)-ATFLG has a pair of perfect flat bands coexisting with two dispersive bands, which is very similar as the ATTLG. Considering the discovery of superconductivity in ATTLG, it strongly hints  that superconductivity can also be detected in the (X+1+Z)-ATFLG. Meanwhile, when the twist angle is smaller than  $\theta=1.70^\circ$, the (X+2+Z)-CTFLG can have two narrow bands with band width less than $5$ meV, which are well separated from other  bands by a gap larger than 30 meV. It  implies that the (X+2+Z)-CTFLG is also an ideal platform to study the correlation effects of the moir\'{e} flat (or narrow) bands. Since that the double twisted moir\'{e} heterostructures have been realized in experiment, the manufacture of  the (X+1+Z)-ATFLG and  the (X+2+Z)-CTFLG should  be not  too challenge. So, our predictions above can be readily tested in future experiments.  

This paper is organized as follows. In Sec. II, we gives the models of the DTFLGs used in our calculations. In Sec. III, the detailed discussions about the calculated moir\'{e} band structures  are given. At last, a summary is given in Sec. IV.

\section{MODEL and Methods}
We first discuss the  continuum model of a general DTFLG, which was given in Ref.~\onlinecite{2021Doubled}. Here, we give a short introduction about the continuum Hamiltonian. 
For a  general (X+Y+Z)-DTFLG, the continuum Hamiltonian  is,
\begin{equation}\label{eq1}
\begin{aligned}
H_{\mathrm{DTFLG}}=\left(
\begin{array}{ccc}
H_1(k_1)&T_{12}(r)&0\\
T^\dag_{12}(r)&H_2(k_2)&T_{23}(r)\\
0&T^\dag_{23}(r)&H_3(k_3)\\
\end{array}
\right)+V.
\end{aligned}
\end{equation}
Here,  $l=1,2,3$ is the index of the vdW layers, which describes the bottom, the middle, and the top vdW layer, respectively. For example, $H_1(k_1)$ is the $k \cdot p$ Hamiltonian of the bottom vdW layer, where $k_l=k-K_{\xi}^{(l)}$  and $K_{\xi}^{(l)}$ denotes the Dirac point of vally $\xi$. 

\begin{figure}[ht]
\centering
\includegraphics[width=7cm]{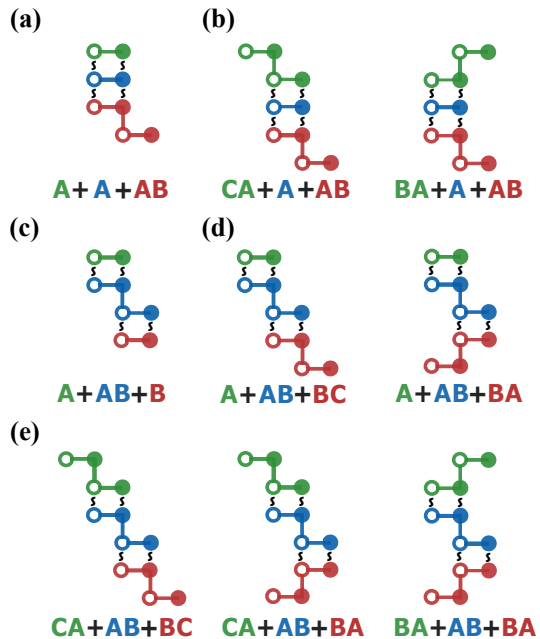}
\caption{Schematic of all the possible arrangements of the DTFLG. (a) is for the (1+1+2)-DTFLG, (b) is for the (2+1+2)-DTFLG, (c) is for the (1+2+1)-DTFLG, (d) is for the (1+2+2)-DTFLG, (e) is for the (2+2+2)-DTFLG. Here, the ATFLG and CTFLG have the same arrangements.  Green, blue and red represent the top, middle and bottom vdW layers, respectively.
}
\label{configuration}
\end{figure}

\begin{figure*}[ht]
\centering
\includegraphics[width=17cm]{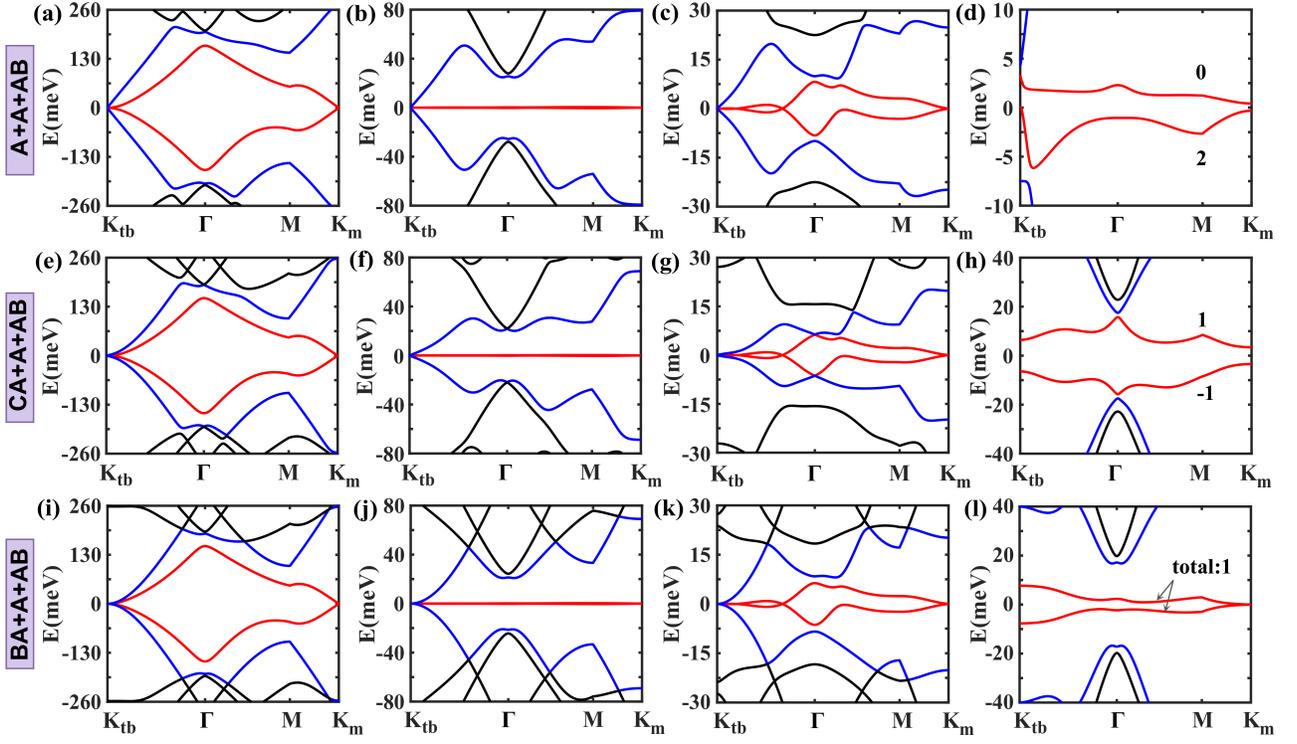}
\caption{The electronic structure of (X+1+Z)-ATFLG. (a-d), (e-h), and (i-l) are the moir\'{e} bands of the (A+A+AB)-, (CA+A+AB)-, and (BA+A+AB)-ATFLG, respectively. (a,~e,~i) is for  $\theta=2.88^{\circ}$; (b,~f,~j) is for $\theta=1.54^{\circ}$; (c,~g,~k) is for $\theta=1.08^{\circ}$. (d,~h,~l) show the influence of electric field at $\theta=1.54^{\circ}$, and the valley Chern numbers are labeled. The electric field:  (d) $\Delta=5$~meV (h) $\Delta=25$~meV and (l) $\Delta=20$~meV. 
}
\label{fig3} 
\end{figure*}

The matrix $T_{ij}$ represents the moir\'{e} interlayer coupling between two adjacent vdW layers, where $i$ and $j$ are layer indexs. 
 $T_{ij}=\sum_{n=0,1,2}T^n_{ij}\cdot{}e^{iq_nr}$, where
\begin{equation}
T_{i j}^{n}=I_{i j} \otimes\left(\begin{array}{cc}
\omega_{\mathrm{AA}} & \omega_{\mathrm{AB}} e^{i \phi_{n}} \\
\omega_{\mathrm{AB}} e^{-i \phi_{n}} & \omega_{\mathrm{AA}}
\end{array}\right).
\end{equation}
Here, $I_{ij}$ is a matrix with only one nonzero matrix element\cite{2021Doubled}. For instance, in a (X+Y+Z)-DTFLG, the only nonzero matrix element of $I_{12}$ is $I_{12}(X,1) = 1$.
 $\phi_{n}=\textrm{sign}(\theta_{ij}) \frac{2n\pi}{3}$, where $\theta_{ij}$ is the twist angle between two adjacent vdW layers. As shown in Fig.~\ref{fig1}, we have two twist angles $\theta_{12}$ and $\theta_{23}$ in a DTFLG, and the sign of the twist angles represents the rotation direction. Thus, in alternate (chiral) twist case, $\theta_{12}$ and $\theta_{23}$ have the opposite (same) sign. Here, we assume that $\theta \equiv |\theta_{12}|=|\theta_{23}|$, otherwise  a moir\'{e} supercell can not be found. It should be noted that the continuous Hamiltonian above is only valid when $\theta$ is small for the chiral twist case\cite{Mora2019}. 
 The other two parameters in $T^n_{ij}$ are  $\omega_{\mathrm{AA}}=0.0797$ eV, and $\omega_{\mathrm{AB}}=0.0975$ eV\cite{PhysRevB.99.235406,koshino2018}. Moreover, in $T_{ij}$ above,  $q_{n}=2 k_{D} \sin \left(\frac{\theta}{2}\right) \exp \left(i \frac{2 n \pi}{3}\right)$, where $k_D$ is the magnitude of the BZ corner wave vector of a single vdW layer. 
 
 In the continuum Hamiltonian \eqref{eq1},  $V$ represents the potential induced by applying a vertical electric field.  Here, we assume that the potential distributes uniformly bewtween layers, and $\Delta$ denotes the potential difference between two adjacent layers.  Take the (1+2+1)-DTFLG for example, $V$ is a diagonal matrix $V=\mathrm{diag} (\frac{3}{2}\Delta\hat{1},\frac{1}{2}\Delta\hat{1},-\frac{1}{2}\Delta\hat{1},-\frac{3}{2}\Delta\hat{1})$,
where  $\hat{1}$ is a $2\times2$ unit matrix. 

\begin{figure*}[ht]
\centering
\includegraphics[width=17cm]{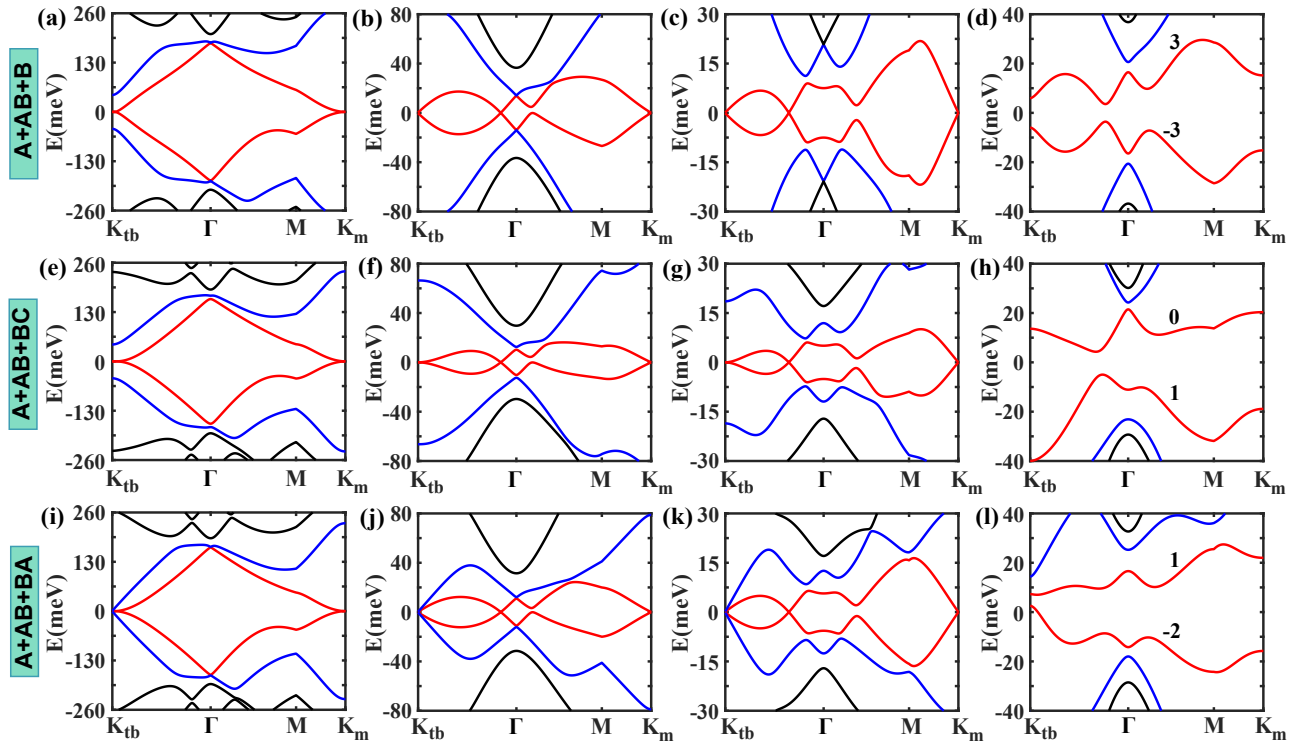}
\caption{The electronic structure of (1+2+Z)-ATFLG. (a-d), (e-h), and (i-l) are the moir\'{e} bands of the (A+AB+B)-, (A+AB+BC)-, and (A+AB+BA)-ATFLG, respectively. (a,~e,~i) is for $\theta=2.88^{\circ}$; (b,~f,~j) is for $\theta=1.54^{\circ}$; (c,~g,~k) is for $\theta=1.08^{\circ}$; (d,~h,~l) show the influence of electric field at $\theta=1.54^\circ$, and the valley Chern numbers are labeled. The electric field: $\Delta=20$~meV.
}
\label{fig4}
\end{figure*}

Now, we discuss the arrangements of the DTFLG.  Actually, with different arrangements of MLG and BLG, we can get various kinds of DTFLG. We denote these arrangements of DTFLG by their stacking configuration when $\theta=0$.  And, for simplicity,  we further require that both the two moir\'{e} interfaces in the DTFLG will restore to a AA configuration when $\theta=0$, which is the most possible case as suggested in the recent experiment\cite{park_tunable_2021}. It is noted that, in the double twisted moir\'{e} heterostructures, the kind of moir\'{e} interface will influence the band structures\cite{lixiaotrilayer,PhysRevB.104.035139,Shin2021,KoshinoprB.101.041112}. Here, we leave the  DTFLGs with other kinds of  moir\'{e} interface to future works.

In Fig.~\ref{configuration}, we show all the possible arrangements of DTFLG that satisfy the requirements above. It is known that in the multilayer graphene, a MLG has three different placements, denoted as A, B, C. For convenience, we always set the MLG (BLG) of the middle vdW layer to be the A (AB) position. Then, we see that the (1+1+2)-DTFLG has only one possible arrangement, \textit{i.e.}, (A+A+AB)-DTFLG as shown in Fig.~\ref{configuration} (a). In contrast, the (2+1+2)-DTFLG has two, namely (CA+A+AB)-DTFLG and (BA+A+AB)-DTFLG, see Fig.~\ref{configuration} (b). Note that, in such two arrangements, the top and bottom vdW layers (BLG) have have different stacking chirality, and will give rise to distinct moir\'{e} band structures. It is quite like the case of the TDBG, where AB+AB and AB+BA are two distinct configurations and have different band structures\cite{PhysRevB.99.235406}. 
 With similar reasons, the (1+2+2)-DTFLG has two different arrangements [Fig.~\ref{configuration} (d)], the (2+2+2)-DTFLG has three ones [(Fig.~\ref{configuration} (e)], while the (1+2+1)-DTFLG has only one [Fig.~\ref{configuration} (c)]. Given the two  twist manners (alternate or chiral twist), we totally consider $18$ kinds of DTFLG in this work.

\section{RESULTS AND DISCUSSIONS}

\subsection{Electronic structure of ATFLG}
We first discuss the moir\'{e} band structures of the alternate twist case. It is known that the ATTLG has a pair of perfect flat bands coexisting with a pair of linear bands at $E_f$. Our calculations show that, once the BLGs are used to build the DTFLG, what kinds of moir\'{e} bands it has strongly depends on that the middle vdW layer is MLG or BLG.  Thus, we have two categories of ATFLGs, \textit{i.e.}~(X+1+Z)-ATFLG and (X+2+Z)-ATFLG. 

The moir\'{e} Brillouin zone (BZ) of the ATFLG is the key factor to understand its moir\'{e} bands, which is shown in Fig.~\ref{fig1} (c).  Since in the alternate twist case $\theta_{12}=-\theta_{23}$, the BZ of the top vdW layer coincides exactly with that of the bottom vdW layer. Thus, the Dirac point $K_{tb}$ of the moir\'{e} BZ corresponds to the Dirac points of the top and bottom vdW layers, while the other Dirac point $K_m$ is from the middle vdW layer.

\begin{figure*}[ht]
\centering
\includegraphics[width=17cm]{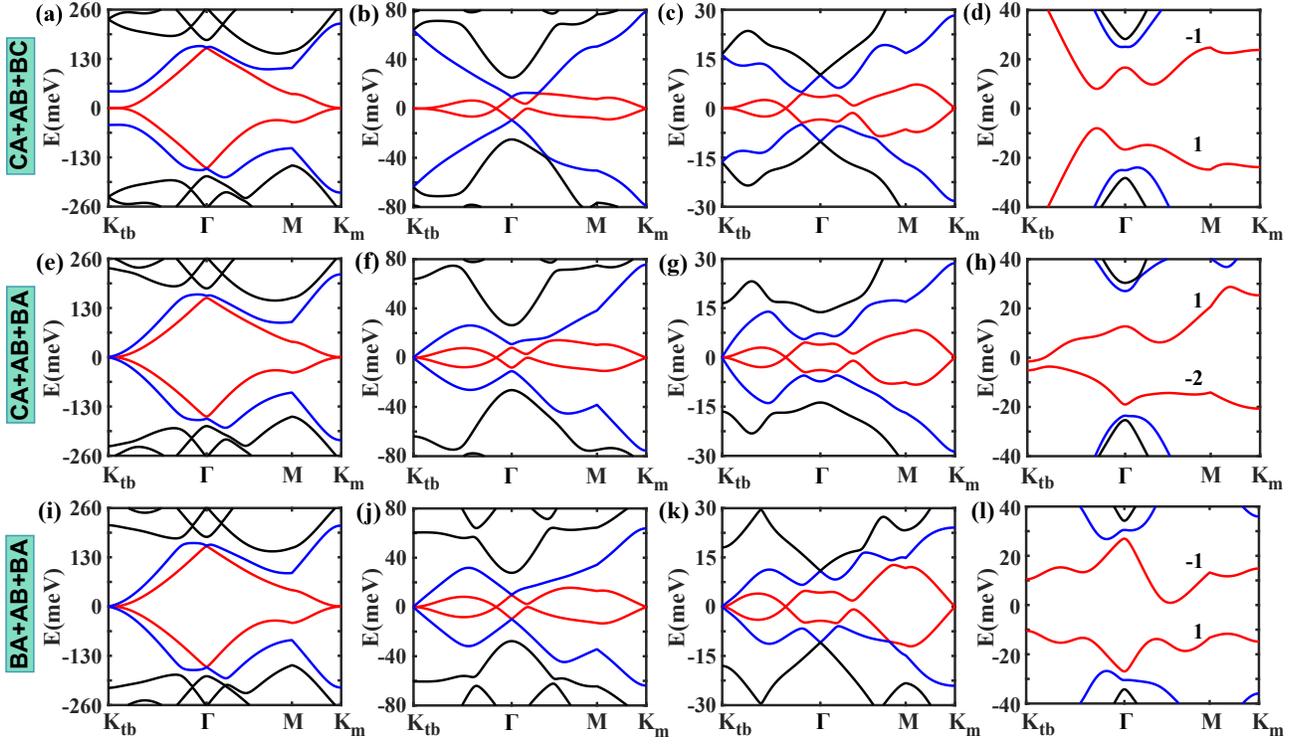}
\caption{The electronic structure of (2+2+2)-ATFLG. (a-d), (e-h), and (i-l) are the moir\'{e} bands of the (CA+AB+BC)-, (CA+AB+BA)-, and (BA+AB+BA)-ATFLG, respectively. (a,~e,~i) is for $\theta=2.88^{\circ}$; (b,~f,~j) is for $\theta=1.54^{\circ}$; (c,~g,~k) is for $\theta=1.08^{\circ}$; (d,~h,~l) show the influence of electric field at $\theta=1.54^{\circ}$, and the valley Chern numbers are labeled.The electric field:  (d) $\Delta=20$~meV (h) $\Delta=20$~meV and (l) $\Delta=40$~meV. 
}
\label{fig5}
\end{figure*}

\subsubsection{(X+1+Z)-ATFLG}
As shown in Fig.~\ref{configuration}, (X+1+Z)-ATFLG has three different arrangements, \textit{i.e.}~A+A+AB,  CA+A+AB and BA+A+AB. The moir\'{e} bands of the three arrangements are plotted in Fig.~\ref{fig3}.


At a large twist angle $\theta=2.88^\circ$, it is clear that all the three arrangements have a pair of linear bands touching at  $K_m$, see Fig.~\ref{fig3} (a), (e) and (i). It is because that the states near $K_m$ mainly result from the middle vdW layer, \textit{i.e.}~the MLG in (X+1+Z)-ATFLG.  Meanwhile, at $K_{tb}$, the A+A+AB arrangement has a pair of $k^2$ bands and a pair of linear bands both touching at $E_f$, see Fig.~\ref{fig3} (a), where the linear bands are from the top vdW layer (MLG, blue lines) and the parabolic bands belong to the bottom vdW layer (BLG, red lines). In comparison,  in the CA+A+AB and BA+A+AB arrangements, the top and bottom vdW layers are both BLG, so that each of them will contribute one pairs of parabolic bands at $K_{tb}$. However, due to the moir\'{e} interlayer hopping, the effective masses of two pairs of $k^2$ bands become different, see Fig.~\ref{fig3} (e) and (i). 

Note that the two linear bands at $K_m$ are hybridized with  one pair of $k^2$ bands at $K_{tb}$ to form a pair of moir\'{e} bands near $E_f$, see the red lines in Fig.~\ref{fig3}. Meanwhile, the other two bands at $K_{tb}$ (blue lines) are connected with high energy bands, which are more dispersive than the red ones. Here, by comparing  Fig.~\ref{fig3} (e) and (i), we  notice  that, though the central two bands have similar shape,  the other moir\'{e} bands of the CA+A+AB and BA+A+AB arrangements are slightly different, which means that the stacking chirality indeed influences the moir\'{e} band structure.   Such differences will be more evident when $\theta$ decreasing, as will shown later. 

\begin{figure*}[ht]
\centering
\includegraphics[width=17cm]{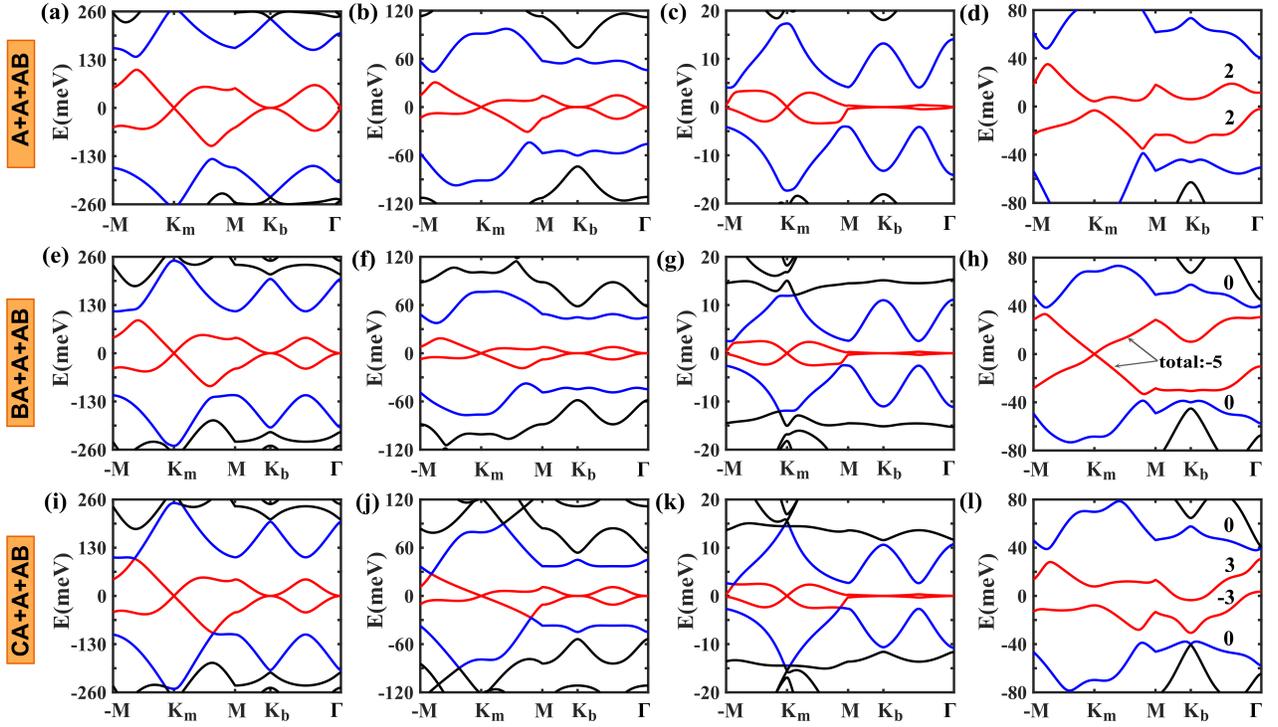}
\caption{The electronic structure of (X+1+Z)-CTFLG. (a-d), (e-h), and (i-l) are the moir\'{e} bands of the (A+A+AB)-, (BA+A+AB)-, and (CA+A+AB)-CTFLG, respectively. (a,~e,~i) is for  $\theta=2.88^{\circ}$; (b,~f,~j) is for $\theta=1.70^{\circ}$; (c,~g,~k) is for  $\theta=0.82^{\circ}$; (d,~h,~l) show the influence of electric field at $\theta=1.70^{\circ}$, and the valley Chern numbers are labeled. The electric field: $\Delta=20$~meV.
}
\label{fig6} 
\end{figure*}

At the magic angle $\theta=1.54^\circ$,  a pair of perfect flat bands appears in all the arrangements of the (X+1+Z)-ATFLG, as shown in  Fig.~\ref{fig3} (b), (f) and (j). The magic angle here is the same as that in the ATTLG, which is larger than the magic angle $\theta=1.08^\circ$ in TBG.  In addition to the two perfect flat bands, the other two bands at  $K_{tb}$ are still dispersive, and remain connected with high energy bands, see the blue lines in Fig.~\ref{fig3} (b), (f) and (j). An  interesting issue is that, in  the A+A+AB arrangement, the perfect flat bands coexist with a pair of linear bands, see Fig.~\ref{fig3} (b). It means that, near $E_f$, the band structures of the (A+A+AB)-ATFLG is exactly the  same as that of the ATTLG, while they have different lattice symmetry.  Due to the very similar band structure, a natural expectation is that superconductivity will also occur in the (A+A+AB)-ATFLG if the lattice symmetry is not the critical factor. In contrast, there are two $k^2$ bands at $K_{tb}$  in the CA+A+AB and BA+A+AB cases, coexisting with the perfect flat bands, see  Fig.~\ref{fig3} (f) and (j). Here, the effective mass of $k^2$ bands in CA+A+AB is larger than that in BA+A+AB.  Because the parabolic bands have finite DOS, the DOS at $E_f$ in the CA+A+AB and BA+A+AB arrangements are larger than that in ATTLG and the A+A+AB case. The enlarged DOS at $E_f$ implies stronger electron correlation effects.
Note that, though the CA+A+AB and BA+A+AB arrangements have similar band structure near $E_f$, their band structures at the magic angle become quite different at high energy.

Decreasing $\theta$ further, the central two perfect flat bands will become dispersive but still have a very narrow band width. For example, the moir\'{e} bands of the (X+1+Z)-ATFLG are plotted in Fig.~\ref{fig3} (c), (g) and (k) with $\theta=1.08^\circ$. Here, the two outer dispersive bands (blue lines) as well as other high energy bands get very close to $E_f$. 

The numerical results above indicate that, though the (X+1+Z)-ATFLGs have different  stacking arrangements, 
 the central two moir\'{e} bands (red lines), \textit{i.e.}~the moir\'{e} flat bands, always have very similar band shape at the same twist angle. However, the outer two dispersive bands (blue lines)  have different behaviours in different arrangements, which is rather obvious at  small twist angle.

Now, we discuss the influence of external electric field. Without external electric field, the degeneracy at the $K_m$ and $K_{tb}$ points are irrelevant to the twist angle. When a vertical electric field is applied, it is not surprising that degeneracy of the $k^2$ bands at the Dirac points are lift, since they are from the BLG and electric field will open a gap at $E_f$.  However, the degeneracy of the linear bands have different behaviours. The cases at the magic angle should be of most interest, which are plotted in Fig.~\ref{fig3} (d), (h) and (l). We see that, in the BA+A+AB arrangement, the linear bands at $K_m$ remain degenerate even in the presence of electric field [Fig.~\ref{fig3} (l)], while such degeneracy are lift in the A+A+AB and BA+A+AB arrangements [Fig.~\ref{fig3} (d) and (h)]. Furthermore, when the electric field is applied, the isolated flat bands here (red lines) have nonzero valley Chern number, as denoted in Fig.~\ref{fig3} (d), (h). And in Fig.~\ref{fig3} (l), though the two flat bands are connected at $K_m$, the total valley Chern number of the central two flat bands are nonzero. The numerical results here  show that, for the moir\'{e} flat bands, different stacking arrangements of the (X+1+Z)-ATFLG have different band topology (valley Chern number).

\begin{figure*}[ht]
\centering
\includegraphics[width=17cm]{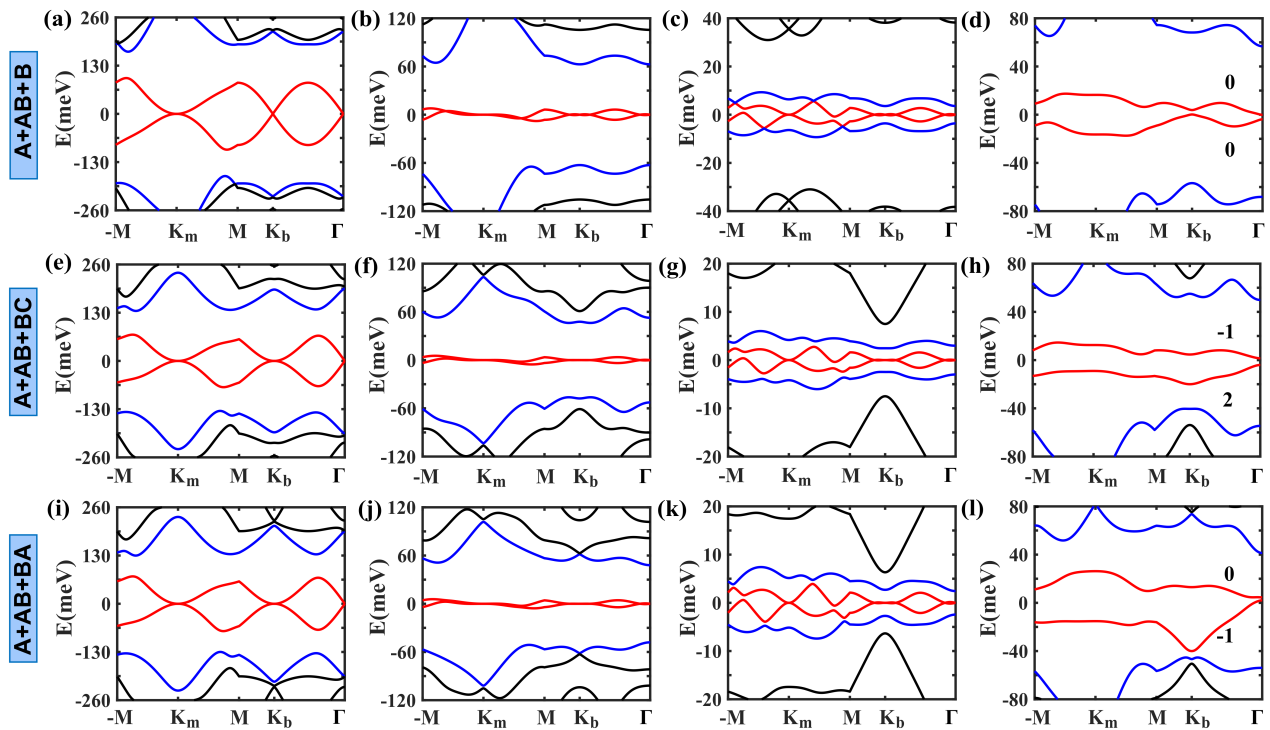}
\caption{The electronic structure of (1+2+Z)-CTFLG. (a-d), (e-h), and (i-l) are the moir\'{e} bands of the (A+AB+B)-, (A+AB+BC)-, and (A+AB+BA)-CTFLG, respectively. (a,~e,~i) is for $\theta=2.88^{\circ}$; (b,~f,~j) is for $\theta=1.7^{\circ}$; (c,~g,~k) is for $\theta=0.82^{\circ}$; (d,~h,~l) show the influence of electric field at $\theta=1.70^{\circ}$, and the valley Chern numbers are labeled. The electric field: (d) $\Delta=20$~meV (h) $\Delta=10$~meV and (l) $\Delta=20$~meV.
}
\label{fig7} 
\end{figure*}

\subsubsection{(X+2+Z)-ATFLG}
When the middle vdW layer becomes a BLG, it will give rise to a completely different moir\'{e} band structures. As mentioned above, we always fix the  middle BLG to be the AB arrangement. Correspondingly, (1+2+1)-ATFLG has only one arrangement, \textit{i.e.}~A+AB+B, as shown in Fig.~\ref{configuration} (c). The (1+2+2)-ATFLG have two different arrangements, A+AB+BC and A+AB+BA, see Fig.~\ref{configuration} (d). As for the (2+2+2)-ATFLG, there are three ones, namely CA+AB+BC, CA+AB+BA and BA+AB+BA. For convenience, we plot the moir\'{e} bands of the (1+2+1)-ATFLG and (1+2+2)-ATFLG in Fig.~\ref{fig4}, while that of the (2+2+2)-aDTFLG in Fig.~\ref{fig5}.

Compared with the (X+1+Z)-ATFLG, the moir\'{e} bands of the (X+2+Z)-ATFLG has two main differences. First, it  is quite clear  that there is no perfect flat bands at any twist angle, as shown in Fig.~\ref{fig4} and Fig.~\ref{fig5}.  The moir\'{e} bands at $E_f$ are always dispersive, though their band widths will decrease when $\theta$ becomes smaller. 

Second,  the  (X+2+Z)-ATFLG have two or four moir\'{e} bands at $E_f$, depending on whether it includes an AB+BA moir\'{e} interface as its component. For example, in the  A+AB+B arrangement [Fig.\ref{fig4} (a), (b), (c)], only two bands (red lines) appears at $E_f$, touching at the $K_{tb}$ point. Meanwhile,  two other bands (blue line) are separated from the $E_f$ by a gap about $40$ meV at $K_{tb}$. In contrast, in the  A+AB+BA arrangement [Fig.~\ref{fig4} (i), (j), (k)], there are four moir\'{e} bands at $E_f$,  touching at $K_{tb}$ point.  According to our numerical results, the rule is that, once the (X+2+Z)-ATFLG has a  AB+BA moir\'{e} interface like that in A+AB+BA, we will find four moir\'{e} bands degenerate at the $K_{tb}$ point. The CA+AB+BA [Fig.~\ref{fig5} (e), (f), (g) ] and BA+AB+BA [Fig.~\ref{fig5} (i), (j), (k)] arrangements belong to this case. 
Otherwise, only two moir\'{e} bands appear at $E_f$, while the other two are separated from $E_f$ by a large gap.  The A+AB+BC [Fig.~\ref{fig4} (e), (f), (g)] and CA+AB+BC arrangements [Fig.~\ref{fig5} (a), (b), (c)] are in this category. 

\begin{figure*}[ht]
\centering
\includegraphics[width=17cm]{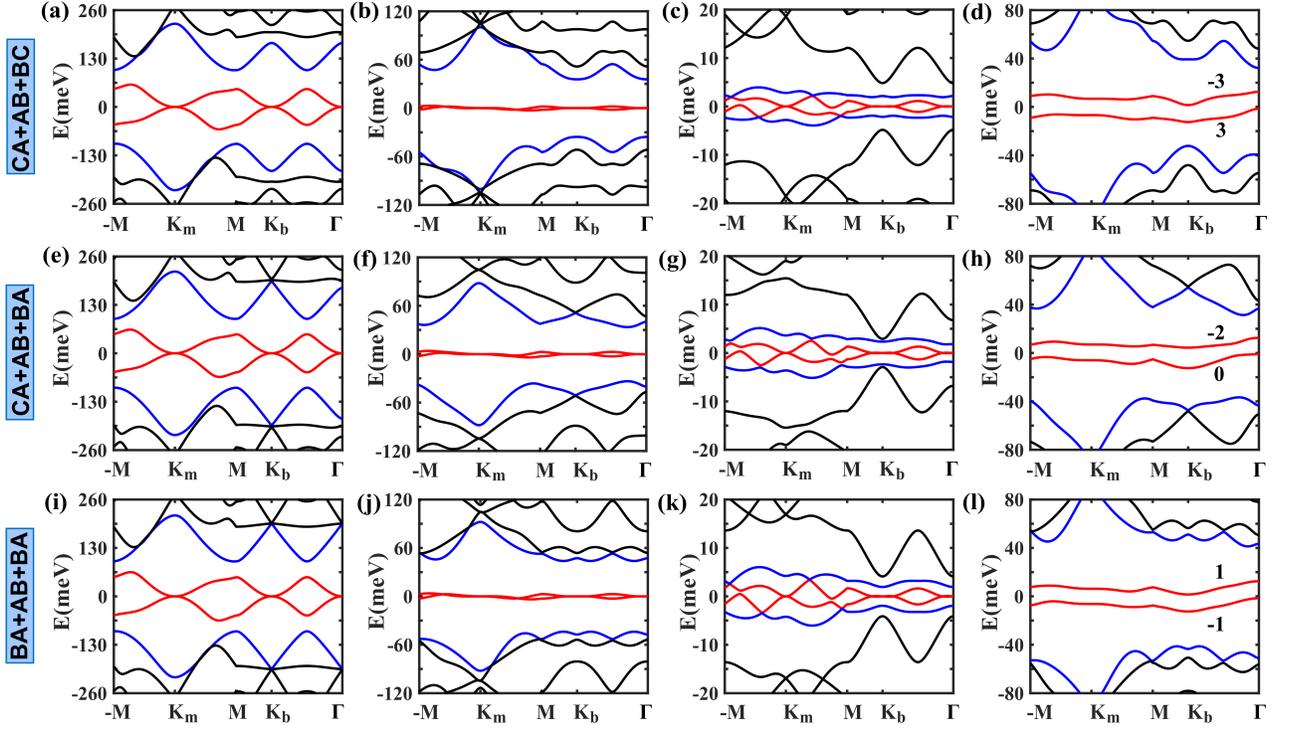}
\caption{The electronic structure of (2+2+2)-CTFLG. (a-d), (e-h), and (i-l) are the Moir\'{e} Bands of the (CA+AB+BC)-, (CA+AB+BA)-, and (BA+AB+BA)-CTFLG, respectively. (a,~e,~i) is for $\theta=2.88^{\circ}$; (b,~f,~j) is for $\theta=1.70^{\circ}$; (c,~g,~k) is for $\theta=0.82^{\circ}$; (d,~h,~l) show the influence of electric field at $\theta=1.70^{\circ}$, and the valley Chern numbers are labeled. The electric field: $\Delta=5$~meV. 
}
\label{fig8} 
\end{figure*}

In Fig.~\ref{fig4} and Fig.~\ref{fig5}, we plot the moir\'{e} bands of various arrangements at three different twist angles, \textit{i.e.}, $\theta=2.88^\circ$ [see (a), (e), (i) of Fig.~\ref{fig4} and Fig.~\ref{fig5}], $\theta=1.54^\circ$  [see (b), (f), (j) of Fig.~\ref{fig4} and Fig.~\ref{fig5}], and $\theta=1.08^\circ$ [see (c), (g), (k) of Fig.~\ref{fig4} and Fig.~\ref{fig5}]. Like the case of (X+1+Z)-ATFLG, the central two moir\'{e} bands (red lines) in different stacking arrangements here always have very similar band shape  at the same twist angle. Meanwhile, the outer two dispersive bands (blue lines) behaves rather differently in the diverse arrangements. 

We then discuss the effects of electric field. For the (X+2+Z)-ATFLG, a vertical electric field can break all the degeneracy at the $K_{tb}$ and $K_m$ points. Thus, the central two moir\'{e} bands can be isolated, and have nonzero valley Chern number, as shown in the subfigures (d), (h) and (l) of Fig.~\ref{fig4} and Fig.~\ref{fig5}. Though the central two moir\'{e} bands are always  dispersive, their band widths are still small if the twist angle is small enough. For example, under an electric field  $\Delta=20$ meV, we can get an isolated narrow band in the (A+AB+BA)-ATFLG at $\theta=1.54^\circ$, where its band width is less than $21$ meV and valley Chern number is $1$, see  Fig.~\ref{fig4} (l). Therefore, with proper twist angle and electric field, it is also possible to observe correlated and topological phenomena in the (X+2+Z)-ATFLG.


\subsection{Electronic structure of CTFLG}
As mentioned above, for CTFLG, our model is valid for the cases where $\theta=\theta_{12}=\theta_{23}$ and $\theta$ is small. We would like to compare the moir\'{e} band structures of CTFLG with that of CTTLG, which has already been realized in experiment. It is known that a CTTLG is a gapless prefect metal, where all the moir\'{e} bands are connected duo to its space symmetry and there is no gap at any energy region. However, if some MLGs in the CTTLG are replaced by BLG, \textit{i.e.}~the so called CTFLG here, it will give rise to a rather distinct moir\'{e} band structures, which are plotted in Fig.~\ref{fig6}, Fig.~\ref{fig7} and Fig.~\ref{fig8}. 

First of all, the most apparent feature of the CTFLG is that it always has a pair of moir\'{e} bands near $E_f$, which are  dispersive at any twist angle. So, we can not get  perfect flat bands in CTFLGs. As will be shown later, except the CA+A+AB arrangement, such two moir\'{e} bands near $E_f$ are isolated from other high energy bands, and  obvious gaps can be found between the central two bands and other high energy bands, when $\theta$ is not too small. This is completely different from the case of CTTLG, where no gaps can be found at all the energy. Here, we first give an intuitive understanding about such two moir\'{e} bands. The moir\'{e} BZ of the CTFLG is given in Fig.~\ref{fig1} (d), where the $\Gamma$ point corresponds to the Dirac point of the top vdW layer, and $K_m$ ($K_b$) belongs to the middle (bottom) vdW layer. In the CTFLGs we discussed here, each of the three Dirac points offers a pair of linear (MLG) or parabolic (BLG) bands, which are hybridized via the moir\'{e} interlayer hopping to form a pair of moir\'{e} bands near $E_f$.

Specifically, the middle vdW layer also plays a key role in determining the moir\'{e} band structure of the CTFLGs. Thus, we still divide the CTFLG into two categories, namely (X+1+Z)-CTFLG and (X+2+Z)-CTFLG, which have distinct moir\'{e} band structures. We then discuss their moir\'{e} bands separately in the following. 

\subsubsection{(X+1+Z)-CTFLG}
The (X+1+Z)-CTFLG and (X+1+Z)-ATFLG have the same stacking arrangements. Thus, as shown in Fig.~\ref{configuration}, the (X+1+Z)-CTFLG has three different arrangements, A+A+AB, CA+A+AB and BA+A+AB.   
The moir\'{e} bands of the (X+1+Z)-CTFLGs are plotted in Fig.~\ref{fig6}, where the first row, the second row and the third row correspond to the A+A+AB, BA+A+AB and CA+A+AB, respectively. 
Here, we consider three different twist angles, \textit{i.e.}~$\theta=2.88^\circ$, $\theta=1.70^\circ$ and $\theta=0.82^\circ$, which are plotted in the first, second and third column, respectively.

\begin{figure}[ht]
\centering
\includegraphics[width=7.5cm]{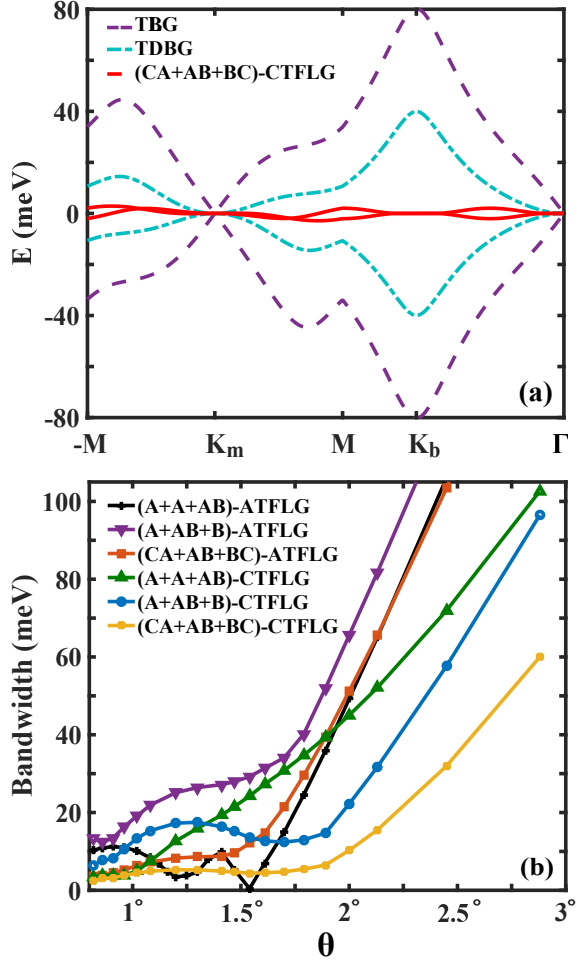}
\caption{(a) The moir\'{e} bands of the TBG, TDBG, and (CA+AB+BC)-CTFLG at $\theta=1.70^\circ$. (b) The band width of the the central moir\'{e} bands at $E_f$ as a function of $\theta$. 
}
\label{fig10} 
\end{figure}

 It is clearly shown that, for the (X+1+Z)-CTFLG,  the pair of moir\'{e} bands at $E_f$ (red lines) in all the three arrangements have similar band shape at the same twist angle. The only difference is that, in the A+A+AB and BA+A+AB arrangements, the central two bands are separated from other high energy bands by an obvious band gap, see Fig.~\ref{fig6} (a), (b), (e) and (f). In contrast, in the CA+A+AB case, such two moir\'{e} bands are connected to the high energy bands (blue lines), see Fig.~\ref{fig6} (i), (j). Note that the case of CA+A+AB arrangement is quite like that of the CTTLG, which actually gives rise a gapless perfect metal without any energy gap in the whole  energy region. In Fig.~\ref{fig6} (c) and (g), we plot the moir\'{e} bands of the A+A+AB and BA+A+AB arrangements   at a very small twist angle, $\theta=0.82^\circ$. In this situation, though the central two bands are still isolated, the gap from the high energy bands becomes very tiny,  which can not be distinguished in experiment. The moir\'{e} bands of the  CA+A+AB arrangement at $\theta=0.82^\circ$ is plotted in Fig.~\ref{fig6} (k) as well. Clearly, when $\theta$ is small enough, the central two moir\'{e} bands of  the  (X+1+Z)-CTFLG  become very narrow, despite they are not perfect flat. It implies that the correlation effects may also occur in the (X+1+Z)-CTFLG as long as $\theta$ is small enough. 

Despite the same band shape, the central two moir\'{e} bands in different arrangements will have diverse topological features, when an external electric field is applied. For example, in the A+A+AB case, an electric field $\Delta=20$ meV will lift the degeneracy at the three Dirac points ($\Gamma$, $K_m$ and $K_b$), so that the central two bands are isolated (red lines). Thus, we can calculate their valley Chern number, as indicated in Fig.~\ref{fig6} (d). In the CA+A+AB arrangement, the case is similar. Though it is gapless without electric field, an electric field can isolated the central two bands as well, see Fig.~\ref{fig6} (l), and the valley Chern numbers are 3 and -3, different from that in the A+A+AB case. The BA+A+AB arrangement behaves differently, see in Fig.~\ref{fig6} (h). The central two bands remain degenerate at the $K_m$ point in this situation, but their total valley Chern number is nonzero. 

\subsubsection{(X+2+Z)-CTFLG}
For convenience, we plot the moir\'{e} bands of the (1+2+Z)-CTFLG in Fig.~\ref{fig7}, and that of the (2+2+2)-CTFLG in Fig.~\ref{fig8}. The corresponding arrangements of the graphene layers are the same as that in ATFLG, as given  in Fig.~\ref{configuration}. In Fig.~\ref{fig7} and Fig.~\ref{fig8}, each row is for one special arrangement, while the first, second and third column corresponds to the twist angle $\theta=2.88^\circ$, $\theta=1.70^\circ$ and $\theta=0.82^\circ$, respectively. 

Similarly, in all the (X+2+Z)-CTFLG, the central two moir\'{e} bands (red lines) have the same band shape at the same twist angle. The band shape  is  different from that of the (X+1+Z)-CTFLG. This feature is shown clearly in  Fig.~\ref{fig7} and Fig.~\ref{fig8}. For all arrangements, the central two bands are separated from the high energy bands by an obvious gap. 

Decreasing the twist angle, the band width of the central two moir\'{e} bands becomes narrow.  An important issue  is that, when $\theta$ approaches about $1.70^\circ$, the band width of the central  bands in (X+2+Z)-CTFLG has already been rather small. For example, at $1.70^\circ$, the band width in the (A+AB+B)-CTFLG is about 11 meV, see Fig.~\ref{fig7} (b). Meanwhile, the value of the (CA+AB+BC)-CTFLG is even less than 5 meV, as shown in Fig.~\ref{fig8} (b).  According to the experience in other moir\'{e} heterostructures like TBG, such band width here is small enough to produce novel correlation effects.   As a comparison, we plot the moir\'{e} bands of the TBG, TDBG, and (CA+AB+BC)-CTFLG at the twist angle $\theta=1.70^\circ$ in Fig.~\ref{fig10} (a). Obviously, at $1.70^\circ$, the band width of the (CA+AB+BC)-CTFLG is much narrower than the TBG and TDBG, which implies that the correlation phenomenon can be observed at a rather large twist angle in the (X+2+Z)-CTFLG.

In Fig.~\ref{fig10} (b), we plot the band width of the central two moir\'{e} bands  as a function of twist angle for various DTFLGs. It indicates that the (CA+AB+BC)-CTFLG always has the narrowest band width, where $\theta=1.70^\circ$ is small enough to  get a narrow band with band width less than 5 meV. Meanwhile, the band width of the (A+AB+B)-CTFLG is not a monotonic function of the twist angle, where the first minimum of the band width (about 11 meV) appears also around the $1.70^\circ$. Therefore, in some sense, $\theta=1.70^\circ$ can be viewed as the ``magic angle'' of the (X+2+Z)-CTFLG, which is larger than the $1.05^\circ$ in the TBG and $1.54^\circ$ in the ATTLG.  Considering the large gap between the central two bands and other high energy bands as mentioned above, we argue that the (X+2+Z)-CTFLG is a promising moir\'{e} platform to study novel correlation effects. Meanwhile, in Fig.~\ref{fig10} (b),  we also plot the band width functions of the (A+A+AB)-ATFLG, (A+AB+B)-ATFLG and (CA+AB+BC)-ATFLG for comparison. The magic angle about $1.54^\circ$ for the (A+A+AB)-ATFLG is shown clearly.


The situations at small twist angle are also very interesting, where we take $\theta=0.82^\circ$ as an  example, see the subfigures (c), (g) and (k) in Fig.~\ref{fig7} and Fig.~\ref{fig8}. 
When $\theta$ is small enough, we find that two other bands (blue lines) get very close to the central two bands. Thus, near $E_f$, we get four nearly flat bands, which will greatly enhance the DOS at $E_f$. The larger DOS implies that correlation effect at $0.82^\circ$ may be much stronger. An ideal case is the (A+AB+B)-CTFLG, as shown in Fig.~\ref{fig7} (c), where the central four narrow bands are well separated from other high energy bands.

Finally, we show the influence of the electric field. In the (X+2+Z)-CTFLG, all the degeneracy at the Dirac points can be lifted by the electric field, as shown in the subfigures (d), (h), (l) of Fig.~\ref{fig7} and Fig.~\ref{fig8}.  We see that, at $\theta=1.70^\circ$, the central two bands are isolated in the presence of  electric field  and their band width are still quite narrow. Meanwhile,  these isolated narrow bands have nonzero the valley Chern number as well. The (A+AB+B)-CTFLG is an exception, where the valley Chern number of the central two bands are always zero, if the electric field $\Delta$ is not larger than 50 meV. 

\section{Summary}
In summary, we theoretically study the moir\'{e} band structures of the DTFLGs, where three MGL or BLG are stacked on top of each other with two twist angles. Our calculations show that the band structures of DTFLG are completely different from the well known ATTLG and CTTLG.  Depending on the relative twist direction and the middle vdW layer, we find that the DTFLG can be classified into four different categories, \textit{i.e.}, (X+1+Z)-ATFLG, (X+2+Y)-ATFLG, (X+1+Z)-CTFLG and (X+2+Z)-(CTFLG),  which have totally different moir\'{e} band  structures. According to the relative placement of the MLG and BLG,  each  category of DTFLG above has one or several different arrangements denoted by their stacking configurations when $\theta=0$, which have very similar but different band structures.   

The moir\'{e} bands of  (X+1+Y)-ATFLG and (X+2+Z)-CTFLG are of special importance. The (X+1+Y)-ATFLG have a pair perfect flat bands at $E_f$ at the magic angle $1.54^\circ$, coexisting with a pair of linear or parabolic bands, which is quite like the case of ATTLG. Most interestingly, at the magic angle, the  moir\'{e} bands of the (A+A+AB)-ATFLG near $E_f$  are exactly the same as the ATTLG, but with different lattice  symmetry. Considering the unconventional superconductivity found in the ATTLG, it is reasonable to expect that superconductivity may also appear in the (A+A+AB)-ATFLG. Meanwhile, due to the additional $k^2$ bands, the DOS at $E_f$ of the (CA+A+AB)- and (BA+A+AB)-ATFLG are even larger than that of the (A+A+AB)-ATFLG at the magic angle, which indicates that they are also ideal platforms to observe moir\'{e} flat bands induced novel correlation effects. 

Although there is no perfect flat bands, we also find a ``magic angle'' about $\theta=1.70^\circ$ for the (X+2+Z)-CTFLG, below which the band width of the central two moir\'{e} bands in  the (CA+AB+BC)-CTFLG becomes less than 5 meV. 
Note that the ``magic angle'' $1.7^\circ$ here is not only larger than the $1.08^\circ$ of the TBG, but also greater than the $1.54^\circ$ of the ATTLG. We thus expect that  the correlation effects in the  (CA+AB+BC)-CTFLG can be found  at a rather large twist angle $\theta=1.70^\circ$ in experiment. For other two kinds of DTFLG, a rather small twist angle is required to get a narrow moir\'{e} bands at $E_f$. 

Our numerical calculations reveal the main features of moir\'{e} band structures of the DTFLGs, and suggest promising directions for future experiments. We argue that the (X+1+Y)-ATFLG and the (CA+AB+BC)-CTFLG are two ideal  moir\'{e} systems  most likely to host exotic correlated phases.  Since the double twist TLGs have already been realized in experiment, we believe that our predictions here can be tested immediately.     

\textit{Note added.}---Recently, we become aware of a preprint which also discuss the moir\'{e} bands of the (2+1+2)-ATFLG\cite{xie_alternating_2021}.

\begin{acknowledgments}
We acknowledge support from the National Natural Science Foundation of China (Grants No.11874160, 12141401,11534001), the National Key Research and Development Program of China (2017YFA0403501), and the Fundamental Research Funds for the Central Universities (HUST: 2017KFYXJJ027). 
\end{acknowledgments}

\bibliography{twist}

\end{document}